\documentclass[aps,prb,twocolumn, preprintnumbers,amsmath,amssymb]{revtex4}
\usepackage{color}
\usepackage{graphicx}
\usepackage{amsmath}
\usepackage{amssymb}
\usepackage{epsfig}
\usepackage{wasysym}
\usepackage{bbm}
\usepackage{multirow}
\renewcommand{\narrowtext}{\begin{multicols}{2} \global\columnwidth20.5pc}
\renewcommand{\v}[1]{{\bf #1}}

\newcommand{\s}{{\sigma}}

\def\be{\begin{eqnarray}}
\def\ee{\end{eqnarray}}

\newcommand{\nn}{\nonumber\\}

\newcommand{\Eq}[1]{Eq.~(\ref{#1})}

\newcommand{\p}{\partial}

\newcommand{\ra}{\rightarrow}

\newcommand{\e}{\epsilon}
\newcommand{\Fig}[1]{Fig.~\ref{#1}}

\begin{document}

\title{Topological insulators and topological non-linear $\s$ models}

\author{Hong Yao}
\affiliation{
Department of Physics,University of California at Berkeley,
Berkeley, CA 94720, USA}
\affiliation{Materials Sciences Division,
Lawrence Berkeley National Laboratory, Berkeley, CA 94720, USA}
\author{Dung-Hai Lee}
\affiliation{
Department of Physics,University of California at Berkeley,
Berkeley, CA 94720, USA}
\affiliation{Materials Sciences Division,
Lawrence Berkeley National Laboratory, Berkeley, CA 94720, USA}

\begin{abstract}
In this paper we link the physics of topological non-linear $\s$ models with that of
Chern-Simons insulators. We show that corresponding to every $2n$-dimensional Chern-Simons
 insulator there is a $(n-1)$-dimensional topological non-linear $\s$ model with the Wess-Zumino-Witten term. Breaking internal symmetry in these non-linear $\s$ models leads to non-linear $\s$ models with the  $\theta$ term. [This is analogous to the dimension reduction leading from $2n$-dimensional Chern-Simons insulators to $(2n-1)$ and $(2n-2)$-dimensional topological insulators protected by discrete symmetries.] The correspondence described in this paper allows one to derive the topological term in a theory involving fermions and order parameters (we shall referred to them as ``fermion-$\s$ models'') when the conventional gradient expansion method fails. We also discuss the quantum number of solitons in topological non-linear $\s$ model and the electromagnetic action of the $(2n-1)$-dimensional topological insulators. Throughout the paper we use a simple model to illustrate how things work.
\end{abstract}

\date{\today}
\maketitle


\section{Introduction}

In the past couple of years the topic of topological band insulators (TBI) protected by time reversal symmetry \cite{kane2005a,kane2005b,bernevig2006a,fu2007a, moore2007,roy2009} has attracted considerable interests. These insulators are described by free fermion Hamiltonians whose band structure are topologically non-trivial.
(Because the mean-field quasiparticle Hamiltonian of a superconductor is also a free fermion theory, the notion of topological band insulator can be simply generalized to include ``topological superconductors''\cite{qi2009,kitaev2009,schnyder2008}.)
Two and three dimensional TBI with Dirac-like edge/surface band(s) have been predicted\cite{bernevig2006b,fu2007b,zhang2009b} and discovered \cite{konig2007,hsieh2008,hsieh2009,chen2009, xia2009}. In addition, a classification of TBI in different spatial dimensions has been achieved\cite{qi2008,kitaev2009,schnyder2008}. A dimensional reduction scheme that allows one to generate lower dimensional TBI from even-dimensional Chern-Simons insulators has been established \cite{qi2008,ryu2009}.

On a different front, the effects of topological terms on the dynamics of Goldstone modes and
the quantum number of solitons and instantons in non-linear $\s$ (NL$\s$) models have a long history, and continue to attract strong interests from the physics community \cite{wilczek}. One of the earliest examples is the charge associated with the solitons of the Peierls-Fr\"{o}hlich order parameter in one-dimensional charge (bond) density wave systems \cite{jackiw1976,SSH,goldstonewilczek,heeger1988}. This arises when the fermion field, which couples to a topologically non-trivial order parameter profile, is integrated out. In terms of the fermion spectrum this amounts to the presence of in-gap eigenstates (or in short zero modes) localized on the soliton.  Another well-known example is the $\theta$-term in the non-linear $\sigma$ model describing one dimensional antiferromagnetic spin chains \cite{haldane1983}. There, depending on whether $\theta=\pi$ or $0$ [Mod($2\pi$)], the half-integer or integer spin chains are gapless or gapped. In condensed matter physics the appreciation of the physical meaning of various topological terms grows as the number of realization of topological NL$\s$ models increases. For example,
recently it is appreciated that, in a NL$\s$ model describing competing orders, one of the effects of the Wess-Zumino-Witten (WZW) term is to make the topological defects of an order parameter to carry the quantum number of its competition order parameter \cite{hu2005,senthil2006,ran2008,grover2008}.

In many-body physics one often encounters Lagrangians which are quadratic in the fermion fields, and in which some fermion bilinears couple to space-time dependent order parameters. This type of action can be viewed as the ``fluctuation-corrected mean-field theory'' of an interacting fermion problem. In a seminal paper Abanov and Wiegmann\cite{abanov2000} gave a systematic derivation of several kinds of topological terms and soliton quantum numbers when the fermions in such fermion-$\s$ models are integrated out. Their method is designed for models where, without the order parameter, the fermions have a {\it gapless Dirac} spectrum.
The order parameters introduce various types of ``mass term'' (i.e.,  fermion bi-linears)  each of which can gap out the Dirac node(s). Under the condition where the number of such mass terms is right, and when the mass matrices anti-commute with the Dirac matrices and with one another, a topological term (which is a pure imaginary term in the Euclidean action) is generated upon integrating out the fermions.
The topological term can fundamentally affect the ground state properties and the low energy dynamics of the order parameters. However, in condensed matter physics gapless systems are not always described by the massless Dirac theory. For these systems the method presented in Ref.~\onlinecite{abanov2000}
may not work and a more general approach is desired.  Moreover, on a conceptual level, it will be very satisfying to understand the relation between two of the areas in topological condensed matter physics: the topological band insulators and the topological NL$\s$ models. This paper serves to fulfill these goals. The main result of this paper can be summarized as follows.

Let $H_0$ be a translationally-invariant free fermion Hamiltonian defined on a $(n-1)$-dimensional lattice system. Now let us add fermion bilinears that couple to a set of $(n+2)$ space-time dependent order parameters $m_a(\vec{x},t)$ ($a=1,...,n+2$), and hence produce the following
Euclidean action:
\be
S&=&\int dt\left[\sum_i \psi^\dagger_{i}\p_t\psi_i\right.\nonumber\\
&&+\left.H_0+\sum_{a=1}^{n+2} \sum_{i,j}m_a(\vec{x},t) \psi^\dagger_i(t)M^a_{ij}\psi_j(t)\label{ag}\right]\ee
where $i,j$ label the lattice sites and $\vec{x}$ is the averaged position of site $i$ and $j$; $\psi^\dagger_i=(\psi^\dagger_{i1},...,\psi^\dagger_{iD})$ are $D$ fermion creation operators on site $i$.  We will focus on local coupling so that the $D\times D$ matrices $M^a_{ij}$ vanish sufficiently fast as sites $i$ and $j$ get further apart. The question we are interested in answering is under what condition a WZW term for the
$m_a$'s will be generated upon integrating out the fermions.
In the case where $H_0$ possesses a gapless Dirac spectrum
the answer is provided by Ref.~[\onlinecite{abanov2000}].
More explicitly, if the low energy and long wavelength physics in the continuum limit is described by the following action
\be
S&=&
 \int dt d^d x\bigg[ \psi^\dag \p_t \psi +\sum_{\alpha=1}^{n-1} \psi^\dag(-i\p_\alpha \Gamma^\alpha)\psi \nonumber\\
&&+\sum_{a=1}^{n+2} m_a(\vec x,t) \psi^\dag M^a \psi \bigg],
\ee
where $\Gamma^\alpha$'s are anti-commuting matrices namely $\{\Gamma^\alpha,\Gamma^\beta\}=2\delta^{\alpha\beta}$ and $M^a$ are constant mass matrices.
The condition for the generation of the WZW term is when the matrices $\Gamma^\alpha$'s and the mass matrices $M^a$'s satisfy
\be
{\rm Tr}\left[ \Gamma^{\alpha_1}\cdots\Gamma^{\alpha_{n-1}}M^{a_1}\cdots M^{a_{n+2}}\right]\propto\epsilon^{\alpha_1\cdots \alpha_{n-1} a_1\cdots a_{n+2}},
\ee
Here $\epsilon$ is a totally anti-symmetric tensor.

Gapless but non-Dirac dispersions often appear in condensed matter systems. For instance, $H_0$ can describe quadratic or high order band touching,  or even possesses a Fermi surface. For these cases under what condition will a WZW term be generated upon integrating out fermions?

In this paper we shall show that if integrating out the fermion field in \Eq{ag} generates a level $k$ WZW term
\be
 S_{WZW}&=&-i\frac{2\pi k}{ {\rm Area}(S^{n+1})}\int_0^1 du\int dt d^{n-1}x\e^{a_1\cdots a_{n+2}}\nn
&&~\times n_{a_1}\p_u n_{a_2}\p_0 n_{a_3}\p_1 n_{a_4}\cdots \p_{n-1}n_{a_{n+2}},
\label{hut}\ee
where $n_a=m_a/\sqrt{\sum_b m_b^2}$, then there exists a finite range of $\lambda$ where the following Bloch Hamiltonian
\be
&&H(k_1,\cdots,k_{n-1};p_1,\cdots,p_{n+1})=H_0(k_1,\cdots,k_{n-1})\nn&&
\qquad+\lambda\sum_{a=1}^{n+2} m_a(p_1,\cdots,p_{n+1})M^a(k_1,\cdots,k_{n-1})
\label{bloch}
\end{eqnarray}
describe a $2n$-dimensional Chern-Simons insulator, so long as the Pontryagin index\cite{pontryagin} $P$ of the map
\be
(p_1,\cdots,p_{n+1})\ra (m_1(\vec{p}),\cdots,m_{n+2}(\vec{p}))\label{pam}
\ee
is non-zero. In \Eq{bloch} $M^a(k_1,\cdots,k_{n-1})$ is the Fourier transform of $M^a_{ij}$.
In \Eq{hut}
an auxiliary coordinate $u$ is introduced so that
\be
&&\hat{n}(t,x_1,\cdots,x_{n-1},u=0) =(0,0,\cdots,1)\nn
&&\hat{n}(t,x_1,\cdots,x_{n-1},u=1) =\hat{n}(t,x_1,\cdots,x_{n-1}).
\ee
The Pontryagin index of the map in \Eq{pam} is given by
\be
P&=&{1\over {\rm Area}(S^{n+1})}\int\prod_{i=1}^{n+1} dp_i \e^{a_1,...,a_{n+1}}\nn
&&\times n_{a_1}\p_{p_1}n_{a_2}\p_{p_2}n_{a_3}...
\p_{p_{n+1}}n_{a_{n+2}};
\label{pont}\ee
and the $n$-th Chern number characterizing the $2n$-dimensional Chern-Simons insulator is given by\cite{qi2008}
\be
C_n={1\over n!2^n(2\pi)^n}\int d^{2n} k \e^{i_1i_2\cdots i_{2n}}\textrm{Tr}[f_{i_1i_2}f_{i_3i_4}\cdots f_{i_{2n-1}i_{2n}}],\nonumber
\!\!\!\!\!\!\!\!\!\!\\
\label{cn}
\ee
in which $f_{ij}$ is the momentum space Berry curvature associated with the occupied bands.
We emphasize that the above result is valid regardless of whether the gapless Hamiltonian $H_0$ is Dirac-like or not. Of course once we derived the WZW term, other topological NL$\s$ model can be obtained by breaking the internal $O(n+2)$ symmetry.
In this way a link between the topological NL$\s$ models and Chern-Simons insulators is established.

Conversely if \Eq{bloch} describes a Chern-Simons insulator for a range of $\lambda$, it is possible to pick an
appropriately bounded order parameter field $\vec{m}(t,\vec{x})$, so that the fermion NL$\s$ model given in \Eq{ag} has a WZW term upon fermion integration. The readers are referred to appendix A for a  proof of the above
correspondence.

Actually, the connection between $2n$-dimensional Chern-Simons insulators and $(n-1)$-dimensional topological NL$\s$ models discussed above is a special  case of the following more general relation. Consider a 2$n$-dimensional Bloch Hamiltonian describing a Chern-Simons insulators with a nonzero $C_n$ for a finite range of $\lambda$ ($0<\lambda<\lambda^\ast)$
\begin{widetext}
\be
H(k_1,\cdots,k_{n-1};p_1,\cdots,p_{n+1}) =H_0(k_1,\cdots,k_{n-1})+V(k_1,\cdots,k_{n-1};\lambda m_1(p_1,...,p_{n+1}),..., \lambda m_{n+2}(p_1,...,p_{n+1})).
\label{2n}
\ee
\end{widetext}
If the map  $(p_1,\cdots,p_{n+1})\to(m_1,\cdots,m_{n+2})$ has a nonzero pontryagin index $P$, then the following steps  generate a  $(n-1)$-dimensional fermion-$\s$ model which yields a level $k=C_n/P$ WZW term upon integrating out the fermions. (i) Perform the Fourier transform with respect to $k_1,\cdots,k_{n-1}$ to rewrite \Eq{2n} in real space. (ii) Replace $m_a(p_1,\cdots,p_{n+1})$ with an appropriately bounded space-time dependent fields $m_a(t,\vec{x})$.

The correspondence discussed above establishes a link between topological NL$\s$ models and Chern-Simons insulators. It also allows one to determine whether integrating out fermion field generates topological terms in the NL$\s$ model describing the order parameter dynamics when the conventional gradient expansion of in Ref. [\onlinecite{abanov2000}] fails. The rigorous proof for the above correspondence is presented in appendix A.

In the rest of the paper we shall use a simple model to illustrate the principal ideas. It turns out that for this  simple model the Abanov-Wiegmann method does not work, consequently the scheme we discussed above is necessary to derive its WZW term. \\

\section{A simple model with quadratic band touching (QBT)}
\begin{figure}[tbp]
\begin{center}
\includegraphics[scale=0.37]
{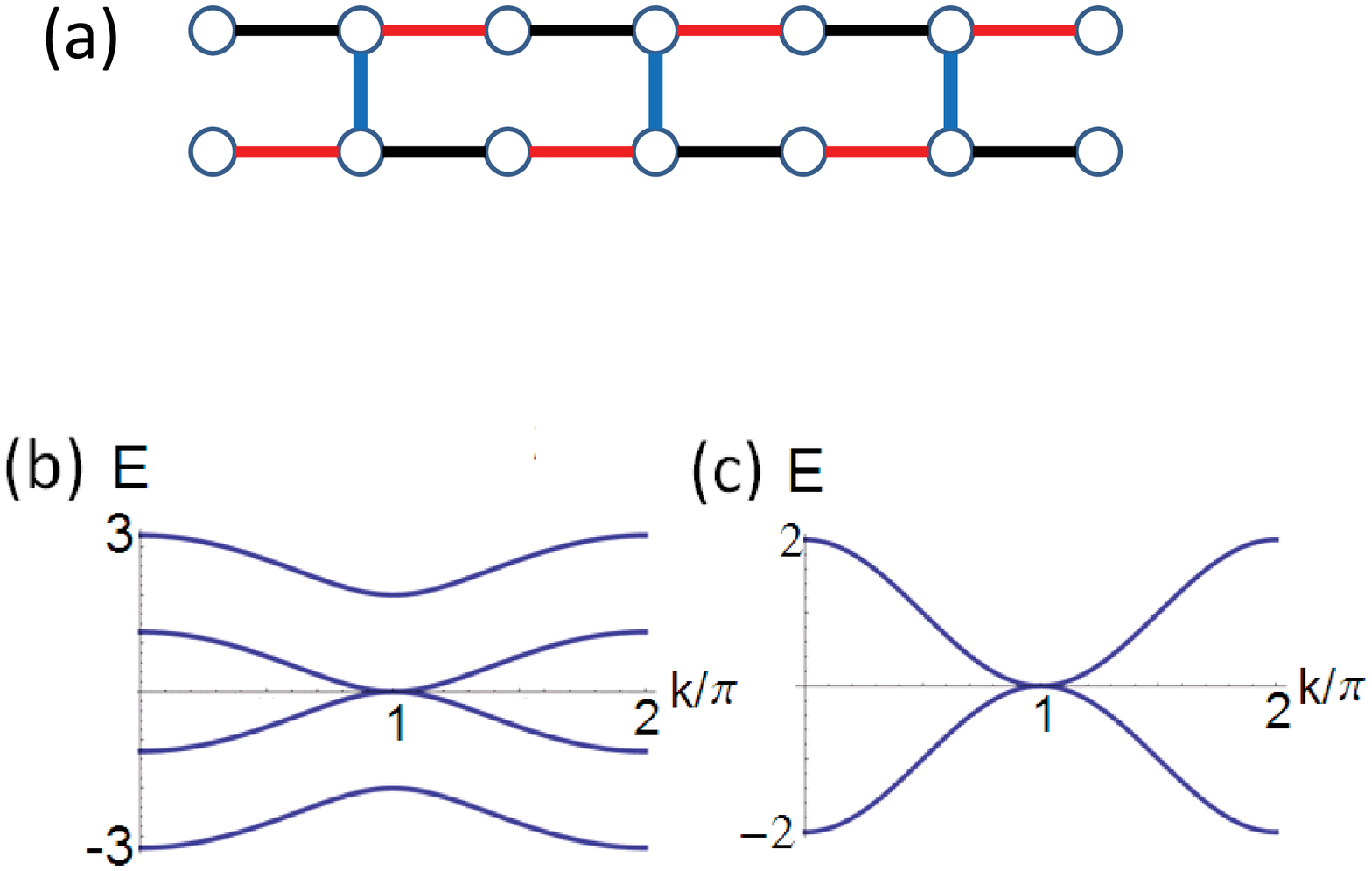}\caption{(color on-line)(a) A schematic representation of the QBT model. Here back and red
indicate strong and weak bonds respectively. The vertical blue bonds are much stronger any other hopping. (b) The dispersion relation for the special case when the back and red bonds have equal strength. (c) The dispersion of the  model in \Eq{qbt} when the strongly hybridizing dimers are eliminated.\label{ladder}}
\end{center}
\end{figure}

As a simple example of a gapless Hamiltonian which has a dispersion different from that of massless Dirac fermions we consider the following one dimensional lattice model (this model is motivated by the spectrum of bilayer graphene). The schematic representation of the model is shown in \Fig{ladder}(a). Here back and red indicate strong and weak bonds respectively. The vertical blue bonds have the strongest hopping. There are four sites per unit cell and the  $4\times 4$ Bloch Hamiltonian is given by
\be
H(k)=
\left(\begin{matrix}
 0 & A & T_{13}(k) & 0 \\
 A & 0 & 0 & T_{24}(k) \\
 T^\ast_{13}(k) & 0 & 0  & 0 \\
 0 & T_{24}^\ast(k) & 0 & 0
\end{matrix}\right)
\label{bloch1d}
\ee
where $T_{13}(k)=(1-\delta t)+(1+\delta t)e^{-ik}$ and $T_{24}(k)=(1+\delta t)+(1-\delta t)e^{-ik}$.
Here $1\pm \delta t$ and $A$ are the strength of black, red and the vertical bonds respectively. When $\delta t=0$ and $A=2$ the dispersion relation is shown in \Fig{ladder}(b). Note that two bands touch {\it quadratically}\cite{chong2008,sun2009}, at $k=\pi$. When $A\gg 1$ the Hamiltonian in \Eq{bloch1d} can be reduced
to an effective $2\times 2$ Hamiltonian by performing second order perturbation theory to remove the strongly hybridizing vertical bonds. The result is given by
\be
-{4\over A}\left(
\begin{array}{cc}
 0  & \left(\cos{k\over 2}+i\delta t\sin{k\over 2}\right)^2 \\
 \left(\cos{k\over 2}-i\delta t \sin{k\over 2}\right)^2 & 0
\end{array}
\right)
\label{reduced}\ee
The corresponding dispersion is shown in part (c) of \Fig{ladder}. In the following we shall multiply the Hamiltonian in \Eq{reduced} by $-{A\over 2}$
and consider the effective model
\be
H(k)&=&2\left(
\begin{array}{cc}
 0  & \left(\cos{k\over 2}+i\delta t\sin{k\over 2}\right)^2 \\
 \left(\cos{k\over 2}-i\delta t \sin{k\over 2}\right)^2 & 0
\end{array}
\right)\nn
&=&\Big[{(1-\delta t^2)+(1+\delta t^2)\cos k}\Big]\tau_1-\left[2\delta t \sin k\right]\tau_2.\nonumber\\\label{qbt}
\ee

One might argue that the QBT point can be viewed as two Dirac points merged together. Hence upon integrating out fermions one can still apply the method of Ref. [\onlinecite{abanov2000}], and sum the results associated with each Dirac point at the end. This is in fact not generally true even though there are cases the conventional gradient expansion work \cite{hongnote}. As the Dirac point merges, the momentum range with linear dispersion vanishes. After the Dirac points merge the length scale above which one can think of the system as two separated Dirac points diverge. It is therefore by no means clear that one can simply sum the results assuming the Dirac points are separated to determine those for the QBT model.

\section{The soliton of the QBT model}
\subsection{The Jackiw-Rebbi soliton}
\begin{figure}[tbp]
\includegraphics[scale=0.37]
{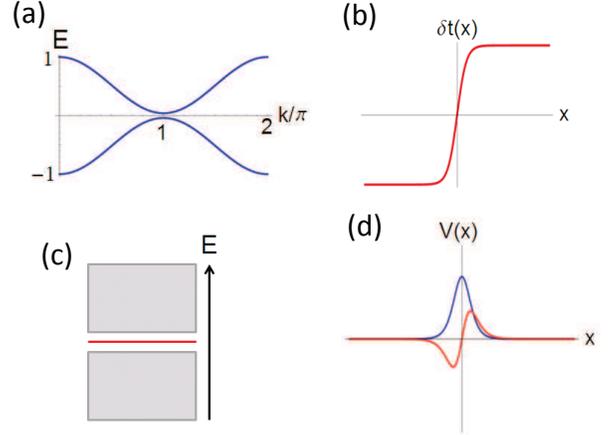}\caption{(color on-line)(a) In the presence of a non-zero $\delta t$ an energy gap opens at the quadratic touching point. Here $\delta t=0.2$ is used. (b) A soliton $\delta t(x)$ profile. (c) Schematic representation of the
energy spectrum associated with a soliton $\delta t(x)$ profiole. The red line marks the position of mid-gap soliton states. (d) The eigenfunctions associated with the zero modes.\label{dw}}
\end{figure}

Setting $\delta t\ne 0$ in \Eq{qbt} opens a gap in the dispersion (\Fig{dw}(a)). In the classic work of Jackiw and Rebbi \cite{jackiw1976} it was shown that in a soliton mass background, the one-dimensional Dirac Hamiltonian exhibits one mid-gap state per soliton. When we let the gap parameter $\delta t$ in \Eq{qbt} to assume a soliton profile (see \Fig{dw}(b)) mid-gap states appear in the energy spectrum (\Fig{dw}(c). However unlike the Dirac Hamiltonian, there are {\it two} mid gap states per soliton, as expected for the QBT. It is worth to point out that on this specific lattice the mid-gap states are true zero modes; they are protected by the ``chiral'' symmetry $\tau_3 H(k) \tau_3 =-H(k)$ in \Eq{qbt}.

\subsection{The zero mode theorem}
In this section we prove the zero mode theorem. In order to allow a comparison with the Jackiw-Rebbi zero mode
theorem\cite{jackiw1976}, we work in the continuum limit. By expanding \Eq{qbt} around $k=\pi$ we obtain
\be
H(q)=\frac12\left(
\begin{array}{cc}
 0 & \left(i 2\delta t-q\right)^2 \\
 \left(-i 2\delta t-q\right)^2 & 0
\end{array}
\right),
\ee
where $q=k-\pi$. When $\delta t$ is position-dependent the above Hamiltonian becomes
\be
&&H=\frac12\left(\begin{matrix}0 &Q^2\\  {Q^\dagger}^2 &0\end{matrix}\right)
\label{ms1}\ee
with
\be Q=i2\delta t(x)+{i}\p_x.\label{ms}\ee Since the third Pauli matrix, $\tau_3$, anticommutes with $H$ in \Eq{ms1}, the zero energy eigenfunctions must be eigenvectors of it. Thus they are either of the form $\begin{pmatrix}0\cr v(x)\end{pmatrix}$ or $\begin{pmatrix}u(x)\cr 0\end{pmatrix}$.

Let us first consider the first possibility
\be
\left(\begin{matrix}0 &Q^2\\  {Q^\dagger}^2 &0\end{matrix}\right)\begin{pmatrix}0\cr v(x)\end{pmatrix}=0.\ee In this case $Q^2 v(x)=0$ and there are two linearly independent solutions, namely,
\be
&&v(x)\sim  e^{-2\int _0^x\delta t(y)dy},~~{\rm and}\nn&&v(x)\sim  x~e^{-2\int _0^x\delta t(y)dy}.\label{sol}\ee
Note that for the $\delta t(x)$ profile shown in \Fig{dw}(b) (where the center of the soliton is at $x=0$) these are decaying solutions as $x\ra\pm\infty$, hence are normalizable.
Note that the first line of \Eq{sol} is the same as the Jackiw-Rebbi soliton solution.
For $\delta t(x)=\delta t_0 \tanh(x/\xi)$ these two
solutions are illustrated as the blue and red curves in \Fig{dw}(d).

The second possibility
\be
\left(\begin{matrix}0 &Q^2\\  {Q^\dagger}^2 &0\end{matrix}\right)\begin{pmatrix}u(x)\cr 0\end{pmatrix}=0,\ee yields the solutions
\be
&&u(x)\sim  e^{2\int _0^x\delta t(y)dy},~~{\rm or}\nn&&u(x)\sim  x~e^{2\int _0^x\delta t(y)dy}.\ee
For the $\delta t(x)$ profile shown in \Fig{dw}(b) these two solution  are not normalizable, hence must be rejected.

\subsection{The spectral flow associated with the Goldstone-Wilczek soliton}
Goldstone and Wilczek\cite{goldstonewilczek} generalized the work of Jackiw and Rebbi\cite{jackiw1976} and considered 
a model with two spatial-dependent mass terms. Viewing the  two mass parameters as the components of a 2-vector, they considered the situation where the mass vector rotates as the position $x$ is varied. In the case where the 1D system is a closed ring there are soliton configurations where the mass vector wind integer number of times as $x$ traverses the ring.
In Ref. \onlinecite{abanov2000} it is shown that for winding number $n$ the soliton charge is $|\Delta Q|=n e$.

\begin{figure}[tbp]
\begin{center}
\includegraphics[scale=0.35]
{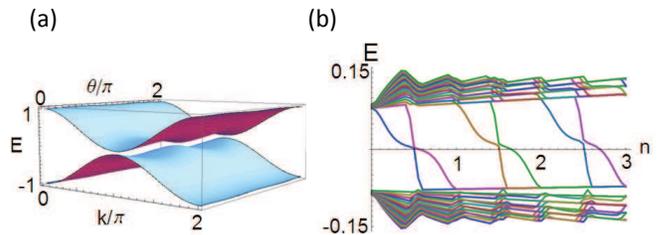}\caption{(color on-line) (a) The energy spectrum of \Eq{GWH}, with $m_1\ra m\cos\theta, m_2\ra m\sin\theta$, as a function of $k$ and $\theta$. Notice that for all values of $\theta$ there is an energy gap in the 1D dispersion $E(k)$. (b) The spectral flow as as function of the winding number $n$. \label{band}}
\end{center}
\end{figure}

 In exactly the same fashion we generalize the QBT model of \Eq{qbt} to include one more mass term. In the situation where both mass terms ($m_1,m_2$) are constant, the Hamiltonian read
\be H(k)=
2\begin{pmatrix}
m_2 & \left(\cos\frac{k}{2} +i m_1 \sin\frac{k}{2}\right)^2\cr
 \left(\cos\frac{k}{2}-i m_1\sin\frac{k}{2}\right)^2 & -m_2
\end{pmatrix}.\label{GWH}
\ee
$m_2$ represents the difference of on-site potential between the two chains.
 First, we let $(m_1,m_2)\ra m (\cos\theta,\sin\theta)$, and plot the energy spectrum of $H(k,\theta)$  as a function of $k$ and $\theta$. The results is shown in \Fig{band}(a) where 
$m$ is chosen to be $0.2$. Clearly, for all values of $\theta$ there is an energy gap. Next, we consider the case where $\theta$ is position-dependent and winds by $2\pi n$ as $x$ traverses the 1D ring. In \Fig{band}(b) We show the spectral flow across $E=0$ as $n$ is varied. (For non-integer $n$ there are spatial discontinuity in the mass terms.)
Clearly in every $2\pi$ cycle of $\theta$,  there are {\it two} energy levels cross $E=0$ from above to below. This is in contrast with the Goldstone-Wilczek model\cite{goldstonewilczek} where only {\it one} level crosses $E=0$ per $2\pi$ cycle.

The above spectral flow is summarized by the following term in the imaginary-time effective action
\be
S_{B}=-{i\over 2\pi}\times 2\int dx A_0 \p_x\theta.\label{eff}\ee The same term with half of the coefficient, i.e., $-{i\over 2\pi}\int dx A_0 \p_x\theta$ was derived first by Goldstone-Wilczek\cite{goldstonewilczek}, and
summarized in Abanov-Wiegmann\cite{abanov2000} as one of the applications of the gradient expansion method.

Thus, to derive the effective action in Eq. (\ref{eff}), one naturally would attempt to generalize the gradient expansion method to the continuum version of the model studied in this section, namely,
\be H=
\begin{pmatrix}
2 m_2(x) & \left({i}\p_x  +i 2 m_1(x)\right)^2\cr
 \left({i}\p_x-i 2 m_1(x)\right)^2 & -2 m_2(x)
\end{pmatrix}.\label{GWH_con}
\ee
\\
\subsection{Beyond the gradient expansion}
It is simple to see
that a direct application of the gradient expansion method employed previously (e.g. in  Ref. \onlinecite{abanov2000}) 
does not work for the model in \Eq{GWH_con}. The problem is rooted in the facts that (1) the two masses $m_1$ and $m_2$ do not enter the Hamiltonian on equal footing. Because of this, it is not possible to find a chiral rotation\cite{fujikawa1979,nagaosa1996} so that locally the mass vector is rotated to a reference value. (2) When we expand  $\left({i}\p_x  +i 2 m_1(x)\right)^2$ the cross term mixes the spatial gradient and $m_1(x)$. In other words, the mass terms couples with the momentum. Nonetheless, we cannot rule out the possibility that some variant of the gradient expansion method can work for the model in \Eq{GWH_con}. Instead of finding a variant of the gradient expansion method, we propose a dimension reduction scheme to derive topological terms in fermionic sigma models. The dimension reduction scheme is qualitatively different from the previous gradient expansion method and is a systematic method applicable to  fermionic sigma models whose $H_0(\vec k)$ can have generic fermion dispersions such as massless Dirac, quadratic touching, or even Fermi surfaces.   
\\
\section{A 2D Chern-Simons insulator and the corresponding 0D topological NL$\s$ models}

In Ref.~\onlinecite{qi2008}, Qi, Hughes, and Zhang put forward a dimension reduction scheme which allows one to
start from Chern-Simons insulators in even dimensions and obtain lower dimensional topological insulators by treating the extra momentum variables as parameters. Here we take their program one step further and show how to obtain topological fermion-NL$\s$ models from dimension reduction.  We will point out some qualitative
differences between our view and that of Qi {\it et al} as we go along. Let us start with the relation between a 2D Chern-Simons insulator and the zero-dimensional NL$\s$ models as a warm up.
\\

\begin{figure}[t]
\begin{center}
\includegraphics[scale=0.35]
{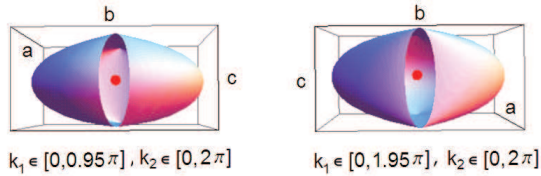}
\vspace{-1in}
\caption{(color on-line) Left panel: the image of $k_1\in[0,0.95\pi],k_2\in [0,2\pi)$. Right panel: the image of $k_1\in[\pi,1.95\pi],k_2\in [0,2\pi)$. We deliberately removed two small intervals in $k_1$, namely, $(0.95\pi,\pi)$ for the left and $(1.95\pi, 2\pi)$ for the right panel, to reveal the origin (red dot) where $H$ is degenerate. \label{abc}}
\end{center}
\end{figure}

\subsection{From the QBT model to a two-dimensional Chern-Simons insulator}

From \Fig{band}(a) we see that by regarding $\theta$ as the momentum in an extra dimension, the bandstrcutre is that of a 2D insulator. What is less obvious is that this insulator is a Chern-Simons  (or Hall) insulator with $\s_{xy}={2\over 2\pi}$, or
equivalently with the first Chern number (TKNN index\cite{tknn}) $C_1=2$. This can be shown by directly computing the first Chern number
\be
C_1={1\over 2\pi}\int_{BZ} d^2k \epsilon^{ij}{\langle u_{\v k}|\p_{k_i} H(\v k)|v_{\v k}\rangle \langle v_{\v k}|\p_{k_j}H(\v k)|u_{\v k}\rangle\over (E_u(\v k)-E_v(\v k))^2}.\label{cn1}\ee
associated with the insulator described by
\begin{widetext}
\be
H(\v k)=2
\begin{pmatrix}m \sin k_2 &\left[\cos\frac{k_1}{2} +im \cos k_2\sin\frac{k_1}{2}\right]^2\cr \left[\cos\frac{k_1}{2}-im \cos k_2 \sin\frac{k_1}{2}\right]^2 & -m \sin(k_2)
\end{pmatrix}.
\label{ch}
\ee
\end{widetext}
In \Eq{cn1} $i,j=1,2$ and $u$ and $v$ labels the occupied and unoccupied bands, respectively. More pictorially we can expand the $2\times 2$ Hamiltonian in \Eq{ch} as the linear combination of three Pauli matrices, and view the
combination coefficients as the coordinates in $R^3$. In this way we have constructed a mapping from the Brillouin zone $k_{1,2}\in[0,2\pi)$ to $R^3$.  In \Fig{abc} we illustrate this mapping. The left panel corresponds to the image of $k_1\in[0,0.95\pi],k_2\in [0,2\pi)$ and right panel is the image of $k_1\in[\pi,1.95\pi],k_2\in [0,2\pi)$. We deliberately remove two small intervals in $k_1$, namely, $(0.95\pi,\pi)$ and $(1.95\pi, 2\pi)$ to reveal the origin (red dot) where $H$ is degenerate.

Clearly the image is a closed surface, and it wraps around the origin {\it twice} which implies $C_1=2$.
If we couple the electromagnetic gauge field to the fermions described by the Hamiltonian in \Eq{ch} and integrate out the fermions, the familiar Chern-Simons action will be obtained
\be
S_{CS_1}=-{i C_1\over 4\pi}\int d^2x dt \e^{\mu\nu\lambda} A_\mu\p_\nu A_\lambda.\label{cs1}\ee
This action summarizes the linear response of the Hall insulator.
\\
\subsection{Explaining the spectral flow of the Goldstone-Wilczek soliton}

Let us imagine the Hall insulator described by \Eq{ch} on a two-torus. We insert a magnetic flux $L_2\phi(x_1)$ through the second hole of the torus and restrict $A_0,A_1$ to depend on {\it $x_1$ and $t$ only}. By substituting $A_2=\phi(x_1)$ and $A_0(x_1,t),A_1(x_1,t)$ into the
Chern-Simons action \Eq{cs1} we obtain
\be
S_{CS_1}\ra -{i C_1\over 2\pi}L_2\int dx_1 dt A_0(x_1,t)\p_1\phi(x_1).\label{gwf}\ee
In obtaining \Eq{gwf} we have taken into account of the fact that the time direction is cyclic.
Since none of $A_{0,1}$ and $\phi$ depend on $x_2$, translation in $x_2$ is a symmetry. As the result, $k_2$ is a good quantum number, and the system decouples into
$L_2$ number of one dimensional subsystems each characterized by a different $k_2$.
The tight binding Hamiltonian of these one dimensional systems are given by
\begin{widetext}
\be
H_{1D}(k_2)&=&\sum_i\psi^\dagger_i\Big[2m\sin(k_2+\phi_i) ~\tau_3+(1-m^2 \cos^2(k_2+\phi_i))~\tau_1\Big]\psi_{i} +\sum_i\bigg[\psi^\dag_{i+1}\Big[{1+m^2\cos^2(k_2+\phi_i)\over 2}~\tau_1\nn &+& im\cos(k_2+\phi_i)~\tau_2\Big]\psi_i+h.c\bigg].
\label{1d}\ee
\end{widetext}
Here $\psi^\dagger_i=(c^\dagger_{1i},c^\dagger_{2i})$, and $\tau_{1,2,3}$ are the Pauli matrices. In the presence of $A_{0,1}(x_1,t)$ each of these one dimensional system contributes an imaginary (Berry phase) effective action $S_{1D}$ which adds up to \Eq{gwf}. Thus
\be
&&\sum_{k_2}S_{1D}[k_2+\phi(x_1);A_\alpha(x_1,t)]\nn&&=-{i C_1\over 2\pi}L_2\int dx_1 dt
A_0(x_1,t)\p_1\phi(x_1)\label{sum}.\ee \Eq{sum} implies
\be &&S_{1D}[k_2+\phi(x_1);A_\alpha(x_1,t)]\nn&&=-i {c(k_2)\over 2\pi}\int dx_1 dt
A_0(x_1,t)\p_1\phi(x_1),
\label{ind}\ee
where
\be \int_0^{2\pi} {dk_2\over 2\pi} c(k_2)={C_1}.\label{adds}\ee

Now we show that $c(k_2)$ is acutally independent of $k_2$. Let us consider the special $k_2=0$ case of \Eq{ind}, and
set $\phi(x_1)=\theta(x_1)+k_2$. Under these condition \Eq{ind} becomes
\be &&S_{1D}[k_2+\theta(x_1);A_\alpha(x_1,t)]\nn&&=-i {c(0)\over 2\pi}\int dx_1 dt
A_0(x_1,t)\p_1\theta(x_1),
\label{ind1}\ee
Compare the above result with the $k_2\ne 0$ case of \Eq{ind} we conclude
$c(k_2)=c(0)$ is independent of $k_2$. Using \Eq{adds} we conclude
$c(k_2)=C_1,~\forall k_2$. Thus
\be
S_{1D}[\phi;A_\alpha]=-{i C_1\over 2\pi}\int dx_1 dt
A_0\p_1\phi.\label{gwe}\ee
\Eq{gwe} is the desired topological term specifying the charge of Goldstone-Wilczek soliton, derived without using the gradient expansion. A similar consideration was given in Ref.\onlinecite{qi2008}, where a {\it spatial constant} but time dependent $k_2$ was considered. The authors argued that the quantum Hall effect in 2D translates into quantized charge pumping as the parameter $k_2$ in the 1D model is varied. In this way they derived $J={C_1\over 2\pi}{d\theta\over dt}$. Then by charge conservation the authors deduced the charge density associated with a spatial dependent $\theta$ to be ${C_1\over 2\pi}{d\theta\over dx_1}.$

The above method deriving the spectral flow of the Goldstone-Wilczek soliton in 1D can be straightforwardly generalized to higher dimensions. To compute the fermion number of a topological soliton in a $n$-dimensional fermion $\s$ model with $n+1$ masses $m_a(t,\vec x)$, $a=1,\cdots,n+1$, we compute the $n$-th Chern number $C_n$ of the corresponding $2n$-dimensional insulator obtained by replacing $m_a(t,x)$ with $m_a(p_1,\cdots,p_n)$, where $p_i$ are momenta in the extra dimensions. When the map $(p_1,\cdots,p_n)\to m_a(p_1,\cdots,p_n)$ has Pontryagin index $P$, the fermion number of the fundamental soliton, i.e., soliton whose real-space Pontryagin index is equal to one,  is $C_n/P$. Since the proof of the above statement is similar to the one for the WZW term, we will not present it explicitly.

\subsection{Deducing the existence of edge states in the QBT model}

In this section we shall deduce informations concerning the edge properties of the QBT model. We start with the 2D Hall insulator (\Eq{ch}) on a torus and adiabatically switch on a spatially independent magnetic flux $L_2\phi$ in, say, the second hole of the torus. Thus $A_2\ra \phi$ and, like in the last section, we shall allow $A_{0,1}$ to depend on $x_1$ and $t$ only. Under these conditions the Chern-Simons effective action becomes
\be
S_{CS_1}\ra {i C_1\phi \over 2\pi}L_2\int dx_1dt \e^{\alpha\beta}\p_\alpha A_\beta.
\label{cs11}\ee
On the other hand since the $x_2$ translation symmetry remains, $k_2$ is a good quantum number. Hence the
action in \Eq{cs11} is the sum of the Berry phase of an assembly of 1D systems, each characterized
by a parameter $k_2$, i.e.,
\be
{i C_1\phi \over 2\pi}L_2\int dx_1dt \e^{\alpha\beta}\p_\alpha A_\beta=\sum_{k_2}S_{1D}[k_2+\phi;A_\alpha].\ee
Compare the above result with that obtained by setting $\phi$ to zero we obtain
\be
&&\sum_{k_2}\left(S_{1D}[k_2+\phi;A_\alpha]-S_{1D}[k_2;A_\alpha]\right)\nn&&=
i C_1{\phi \over 2\pi/L_2}\int dx_1dt \e^{\alpha\beta}\p_\alpha A_\beta.\label{sr}\ee

\begin{figure}[tbp]
\begin{center}
\includegraphics[scale=0.3]
{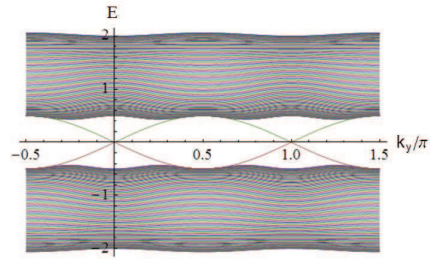}
\caption{(color on-line) The energy spectrum of the 2D Hall insulator with open boundaries in the $x_1$ direction. Vertical cut of this spectrum at fixed $k_2$ yields the energy spectrum of the 1D model associated with parameter $k_2$. The vertical cut at $k_2=0$ gives the energy spectrum of the open-boundary QBT model. }
\label{sb}
\end{center}
\end{figure}

The right hand side of \Eq{sr} is a total derivative, and vanishes in the absence of boundary. Since the imaginary time dimension is cyclic, the only chance for it to be non-zero is when the system is open in the $x_1$ direction. Under that condition \Eq{sr} becomes
\be &&\sum_{k_2}\left(S_{1D}[k_2+\phi;A_\alpha]-S_{1D}[k_2;A_\alpha]\right)\nn&&=-{i C_1}{\phi\over 2\pi/L_2}\int dt \left(A_0(L_1,t)-A_0(0,t)\right)\label{cb}\ee
Perform $i\delta/\delta A_0(L_1,t)$ and $i\delta/\delta A_0(0,t)$  of the right hand side we obtain the extra charge localized on the right and left ends:
\be
&&\sum_{k_2}\Delta Q_{\rm right}(k_2)= C_1{\phi\over 2\pi/L_2},~~{\rm and}\nn&&\sum_{k_2}\Delta Q_{\rm left}(k_2)= -C_1{\phi\over 2\pi/L_2}.\label{qqq}\ee
Here $\Delta Q_{\rm right}(k_2)$ and $\Delta Q_{\rm left}(k_2)$ are the edge state occupation (or change of polarization\cite{king1993theory})
of the adiabatically evolved ($k_2\ra k_2+\phi$) 1D system minus that of the original ($k_2$) system.
In \Fig{sb} we show the energy spectrum of the 2D Hall insulator with open boundary condition in the $x_1$ direction. A vertical cut of this spectrum at fixed $k_2$ yields the energy spectrum of the 1D model associated with parameter $k_2$. In particular, the vertical cut at $k_2=0$ gives the energy spectrum of the QBT model under open boundary condition. Clearly the QBT model possesses zero energy edge states. From the degeneracy and wavefunctions of the
$E=0$ eigenstates we deduce that there is one edge state localized at each end of the system. According to \Fig{sb} in addition to $k_2=0$, the $k_2=\pi$ system also possesses zero energy edge states. This is not so surprising because $k_2=0\ra k_2=\pi$ corresponds to reversing the sign of $\delta t$ in the QBT model. By calculating the eigenfunctions associated with the in-gap band we find that
the upward ($dE(k_2)/dk_2>0$) dispersing  branches near both $k_2=0$ and $\pi$ are associated with the right edge, while the downward dispersing ones are associated with the left edge. This is also not surprising because a $C_1=2$ Hall insulator should have two co-moving chiral edge branches at each edge.

\begin{figure}[tbp]
\begin{center}
\includegraphics[scale=0.35]
{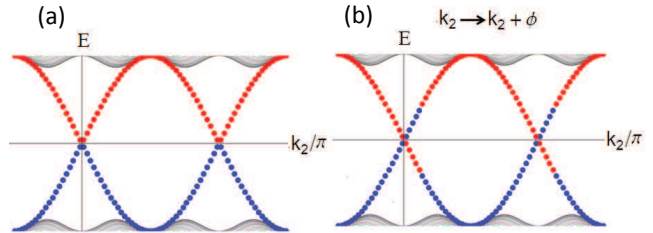}
\vspace{-1in}
\caption{(color on-line) (a) The occupation of the edge states. Blue dots represent occupied states and the red dots
mark the unoccupied states. (b) The occupation after the adiabatic change $k_2\ra k_2+\phi$.  \label{adia}}
\end{center}
\end{figure}

In \Fig{adia} we zoom in on the in-gap band of \Fig{sb}. Here the blue and red dots represent occupied and empty $k_2$ ($=2\pi n/L_2)$ states respectively. Panel (a) is the occupation before the adiabatic evolution and
panel (b) is that after the adiabatic evolution.  For the upward dispersing branch $k_2\ra k_2+\phi$ induces an extra number ($={\phi\over 2\pi/L_2}$) of occupied $k_2$ values near each Dirac point ($k_2=0,\pi$). Similarly, for the downward dispersing branch there are exactly the same number of extra empty $k_2$ values.  As a result, the $C_1$ in \Eq{qqq} counts the number of Dirac points in \Fig{sb}. Equivalently it counts the number of times the edge states cross Fermi energy ($E=0$) as the parameter $k_2$ of the 1D Hamiltonian sweeps from $0$ to $2\pi$. (Note that a crossing from below $E=0$ to above  is counted as $+1$ while the reverse is counted as $-1$.) Of course, the fact that when a parameter of a 1D system is varied, charges flow from one end of the system to the other is the charge pumping effect\cite{thouless1983,qi2008}.

\subsection{The WZW term in $d=0$}
\label{wzw0}

\begin{figure}[t]
\begin{center}
\includegraphics[scale=0.35]{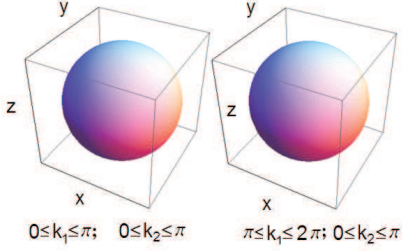}\vspace{-0.5in}
\caption{(color on-line) (a) The map $(k_1,k_2)\ra \hat{n}(k_1,k_2)$, where $k_{1,2}\in [0,2\pi)$.}
\label{map}
\end{center}
\end{figure}

This time we start from the 2D Hall insulator on a torus, and insert fluxes $L_1\theta_1(t)$ and $L_2\theta_2(t)$ through both holes and fix the gauge so that $A_0=0$, $A_1\ra\theta_1(t)$, and $A_2\ra\theta_2(t)$. 
Substitute these changes into the Chern-Simons action (\Eq{cs1}) we obtain
\be
S_{CS_1}\ra S_{CS_1}={i C_1\over 4\pi}L_1L_2\int dt~\e^{ab}\theta_a(t)\p_0\theta_b(t),\label{tcs}\ee where $a,b=1,2$.
Now let us introduce an auxiliary coordinate $u\in [0,1]$ so that $\theta_{1,2}(t,u)$ smoothly interpolate between
$\theta_{1,2}(t,u=0)=0,~{\rm and~}\theta_{1,2}(t,u=1)=\theta_{1,2}(t).$ Then, \Eq{cs1} can be rewritten as
\be
&&S_{CS_1}\nn
&&=-{i C_1\over 4\pi}L_1L_2\int dt\int_0^1 du~\e^{u\nu}\e^{ab}\p_u \Big(\theta_a(t,u)\p_\nu\theta_b(t,u)\Big),\nn
&&=-{i C_1\over 4\pi}L_1L_2\int dt\int_0^1 du~\e^{\mu\nu}\e^{ab}\p_\mu \Big(\theta_a(t,u)\p_\nu\theta_b(t,u)\Big),\nn
&&=-{i C_1 \over 4\pi}L_1L_2\int_0^1 du \int dt~\e^{\mu\nu}\e^{ab}\p_\mu \theta_a(t,u)\p_\nu\theta_b(t,u),\nn
\ee
where $\mu=0$ labels the time and $\mu=1$ labels the $u$ coordinate. In the derivation above, we employed the fact that the only boundary is in the $u$ direction so that integral of a total derivative can be turned into a boundary integral along the $u$-direction.

Since $\theta_1$ and $\theta_2$ do not depend on the spatial coordinates $x_1$ and $x_2$, $k_1$ and $k_2$ remain as good quantum numbers. Thus \Eq{tcs}  should be the sum of the Berry's phase generated by  $L_1 L_2$ number of zero-dimensional systems. Each of this system is parameterized by $k_1$ and $k_2$ and is governed by the following  time-dependent Hamiltonian
\be
H(t;k_{1,2})
=\vec B(t,k_{1,2})\cdot\vec{\tau},
\ee
where the ``effective'' magnetic field $\vec B(t,k_{1,2})$ is given by
\be
B_x(t,k_{1,2})&=&
{1-m^2\cos^2(k_2+\theta_2(t))}\nn &&\!\!\!\!\!+[{1+m^2\cos^2(k_2+\theta_2(t))}]\cos(k_1+\theta_1(t)),\nn B_y(t,k_{1,2})&=&-2m\sin(k_1+\theta_1(t))\cos(k_2+\theta_2(t)),\nn B_z(t,k_{1,2})&=&2m\sin(k_2+\theta_2(t)).\label{h00}
\ee
The three components of the effective magnetic field are just the three masses of the zero-dimensional systems labeled by $k_{1,2}$.
The sum of the Berry's phase $S_B [k_1+\theta_1(t),k_2+\theta_2(t)]$ associated with each $H(t;k_{1,2})$ is equal to $S_{CS_1}$, namely
\be
S_{CS_1}=\sum_{k_1,k_2} S_B [k_1+\theta_1(t),k_2+\theta_2(t)].
\label{tsr}
\ee

Because the following identity which computes the first Chern number\cite{volovik} $C_1$ as well as the Pontryagin index $P$ of the map $(k_1,k_2)\to \hat n\equiv \vec B(\theta_{1,2}(t,u)+k_{1,2})/|\vec B(\theta_{1,2}(t,u)+k_{1,2})|$
\begin{widetext}
\be
C_1=P=\frac1{4\pi}\int {d^2k}\e^{abc}n_a(k_{1,2}+\theta_{1,2}(t,u)){\p\over\p k_1}n_b(k_{1,2}+\theta_{1,2}(t,u)){\p\over\p k_2} n_c(k_{1,2}+\theta_{1,2}(t,u)),
\label{wrap}
\ee
is true regardless of the values of $\theta_1(t,u)$ and $\theta_2(t,u)$ (see \Fig{map}),
we will show that $S_B[k_1+\theta_1(t,u),k_2+\theta_2(t,u)]$ is a WZW term of a spin-1/2 system. To do so, we rewrite $S_{CS_1}$ as follows:
\be
S_{CS_1}&=&-{i C_1\over 4\pi}L_1L_2\int dt\int_0^1 du~\e^{\mu\nu}\e^{\alpha\beta}\p_\mu\theta_\alpha(t,u) \p_\nu\theta_\beta(t,u),
\nonumber\\
&=&-i\frac{C_1/P}{16\pi^2}L_1L_2\int dt\int_0^1 du \int d^2 k ~\epsilon^{\mu\nu} \epsilon^{\alpha\beta}[ \p_\mu \theta_\alpha(t,u) \p_\nu\theta_\beta(t,u)]\nn
&&\qquad\qquad\qquad\qquad\qquad\times\Big[\epsilon^{abc} n_a(k_{1,2}+\theta_{1,2}(t,u)) \p_{k_1} n_b(k_{1,2}+\theta_{1,2}(t,u))\p_{k_2} n_c(k_{1,2}+\theta_{1,2}(t,u)) \Big], \\
&=&-i\frac{C_1/P}{16\pi^2}L_1L_2\int dt\int_0^1 du \int d^2 k ~\epsilon^{\mu\nu} \epsilon^{\alpha\beta}[ \p_\mu \theta_\alpha(t,u) \p_\nu\theta_\beta(t,u)]\nn
&&\qquad\qquad\qquad\qquad\qquad\times\Big[\epsilon^{abc} n_a(k_{1,2}+\theta_{1,2}(t,u)) \p_{\theta_1} n_b(k_{1,2}+\theta_{1,2}(t,u))\p_{\theta_2} n_c(k_{1,2}+\theta_{1,2}(t,u)) \Big],\label{46} \\
&=&\sum_{k_1,k_2}\left[-i\frac{C_1/P}{2}\int dt\int_0^1 du ~\epsilon^{abc} n_a(k_{1,2}+\theta_{1,2}(t,u))\p_t n_b(k_{1,2}+\theta_{1,2}(t,u)) \p_u n_c(k_{1,2}+\theta_{1,2}(t,u))\right],\label{47}
\ee
\end{widetext}
where we employed the chain rule of differentiation from \Eq{46} to \Eq{47}. Because all zero-dimensional subsystems can be adiabatically connected without closing the gap, each of them is then characterized by the same WZW term with the same level. Since $S_{CS_1}=\sum_{k_1,k_2} S_B[k_1+\theta_1(t),k_2+\theta_2(t)]$ and $C_1/P=1$, we conclude that
\be
S_B[\hat n(t)]=-i\frac{1}{2}\int dt \int_0^1 du ~\epsilon^{abc} n_a\p_t n_b \p_u n_c,
\label{wzs}
\ee
which is nothing but the well-known Berry phase or WZW term describing a spin-1/2 in a time-dependent magnetic field, as expected !

\subsection{The $\theta$ term in $d=0$}

The last subsection establishes the fact that the Berry phase (WZW term) associated with the Hamiltonian
$H(t)=\hat{n}(t)\cdot\vec{\tau}$ is \be S_{B}=-{i\over 2}\int dt\int_0^1 du~\e^{abc}n_a\p_t n_b\p_u n_c,\label{sbb}\ee
where $\hat{n}(t,0)=(1,0,0)$ and $\hat{n}(t,1)=\hat{n}(t).$
Physically $S_B$ measures half of the solid angle sustained by  $\hat{n}(t,u)$ at $u=1$  (i.e. the physical $\hat{n}$ trajectory) and the north pole ($\hat{n}(t,u=0)$). For the case where
\be
\hat{n}(t,u=1)=(\sin\theta_0\cos\phi(t),\sin\theta_0\sin\phi(t),\cos\theta_0)\ee \Eq{sbb} becomes
\be
S_B\ra -i\Big({1-\cos\theta_0\over 2}\Big)\int dt {d\phi\over dt}.
\ee
This is the $\theta$-term.

The above is a simple example ($n=1$) of the correspondence between a $2n$-dimensional Chern-Simons insulator and $(n-1)$-dimensional topological NL$\s$ models. This serves as a warm up exercise for the following discussions of the correspondence between a 4D ($n=2$) Chern-Simons insulator and NL$\s$ models in $1$D.

\section{A 4D Chern-Simons insulator and the corresponding 1D topological NL$\s$ models}

In a one dimensional chain (whose gapless excitations near the right and left Fermi points are described by a massless Dirac theory) the dimerization and the antiferromagnetic Neel order parameter form a four vector. The quantum NL$\s$ model describing this ``super-vector'' possesses a topological, WZW, term\cite{senthil2006}. Among others, this means the defects of one order carry the quantum number of the other order. Indeed, in the presence of spin the dimerization  domain wall has two (one for each spin species) zero modes. Among the four ways of occupying them,  two lead to charge-neutral spin 1/2 solitons, and the other two lead to charge $\pm 1$ spin zero solitons. The condensation of the former (which is favored by repulsive interactions) leads to the (algebraic) antiferromagnetic order and the condensation of the latter (which is favored by attractive interactions) leads to (algebraic) charge density wave/superconducting order. The former scenario is encapsulated by the dimerization-antiferromagnetic WZW term discussed above.

In the presence of spin, the QBT model described in section I, has four zero modes per dimerization domain wall (two for each spin). There are totally 16 possible ways to occupy them. Two give rise to charge $\pm 2$ spin zero, eight lead to charge $\pm 1$ spin 1/2, three give charge 0 spin 1, and another three give charge 0 spin 0 solitons. Therefore in this case the destruction of the dimerization order by condescension of domain walls can lead to a large number of possible states. In this section we will focus on the condensation of spin-1 charge neutral solitons. This should lead to spin-1 antiferromagnetic state which has a Haldane gap.  The purpose of the following subsections is to study the WZW term associated with the transition between the dimerized phase and the Haldane phase. In addition, we shall also discuss the edge states of the spin-1 chain, the $\theta$ term and their relation to the physics of 3D TBIs.
\\
\subsection{The QBT model and a 4D Chern-Simons insulator}

In the following we shall find that the $d=1$ topological NL$\s$ models discussed above is related to a Chern-Simons (or quantum Hall) insulator in $d=4$ dimensions, which is constructed in Ref. [\onlinecite{zhang2001four}]. This is the $n=2$ example of the correspondence discussed in the introductory section.
The Hamiltonian for the 4D Chern-Simons insulator is defined as follow:
\be
&&H(k_1,k_2,k_3,k_4)=-\big[(1-m_4^2) + (1+m_4^2)\cos(k_4)\big]\Gamma_5\nn
&&\qquad-m_1\Gamma_1 -m_2\Gamma_2 -m_3\Gamma_3-
  2 m_4 \sin(k_4)  \Gamma_4,
  \label{4dh}\ee
where \be
m_1&=&\sin k_1,\nn
m_2&=&\sin k_2,\nn
m_3&=&\sin k_3, \nn
m_4&=&m+\cos k_1+\cos k_2+\cos k_3,\label{dst}\ee  and
 \be
\Gamma_{1,2,3}=\tau_1\otimes\sigma_{1,2,3}, ~\Gamma_4=\tau_2\otimes I,~\Gamma_5=\tau_3\otimes I
\ee
Here $I$ is $2\times 2$ the identity matrix. The Pauli matrices $\tau$ and $\sigma$ have sublattices (orbitals) and spin degrees of freedom respectively. By setting $k_{1,2,3}$ to zero one can see that the Hamiltonian in \Eq{4dh} and \Eq{dst} reduces to the spinful version of \Eq{qbt}, after a $90^o$ rotation around the $\tau_2$ axis. For, e.g., $m=-2.5$, straightforward computation shows that the above Hamiltonian has a non-zero second Chern number (also see Ref. \onlinecite{qi2008}),  $C_2=2$, where
\be C_2=\frac{1}{4\pi ^2}\int d^4k\text{Tr}\left[f_{12}f_{34}-f_{13}f_{24}+f_{14}f_{23}\right]
\ee
with the momentum space Berry curvature given by
\be
&&f_{\mu \nu}(\v k)=\frac{-i}{\left(E_G(\v k)-E_E(\v k)\right)^2}\\
&&\times\Big( P_G(\v k).\partial _{k_\mu }H(\v k).P_E(\v k).\partial _{k_\nu }H(\v k).P_G(\v k)-\mu\leftrightarrow\nu\Big).\nonumber\label{fmn}\ee
Here we note that because the Hamiltonian is constructed out of $\Gamma$ matrices, its eigenvalues have the form
$(-E,-E,+E,+E)$. In \Eq{fmn} $E_G$ is the energy of the two occupied levels and $E_E$ is that of two unoccupied levels, and the projection operators $P_G,P_E$ are defined as follows
\be
&&P_G(\v k)=\sum_{\alpha={\rm occupied}}|u_{\alpha\v k}\rangle\langle u_{\alpha \v k}|\nn&&
P_E(\v k)=\sum_{\beta={\rm empty}}|u_{\beta\v k}\rangle\langle u_{\beta \v k}|. \ee
After integrating out the fermions, the effective gauge action associated with such a Chern-Simons insulator is\cite{qi2008}
\be
S_{CS_2}=-i{C_2\over 24\pi^2}\int dt d^4x \e^{\mu\nu\lambda\rho\s}A_\mu\p_\nu A_{\lambda}\p_\rho A_{\s}.\label{cs2}\ee
\Eq{4dh} and \Eq{cs2} play the roles of \Eq{ch} and \Eq{cs1} in the following discussions.\\

\subsection{The existence of surface states in 3D TBIs}

We start with the Chern-Simons insulator described by \Eq{4dh} and \Eq{dst} on a real-space 4-dimensional lattice with periodic boundary conditions (hence form a 4D torus). Let us insert flux $L_4\phi$ in the fourth hole (hence $A_4\ra\phi$) and restrict $A_{0,1,2,3}$ to depends on only $t$ and $x_{1,2,3}$. Under this condition \Eq{cs2} becomes
 \be S_{\text{eff}}=-i\frac{C_2\phi }{8 \pi ^2}L_4\int d^3x dt \epsilon ^{\nu \lambda \rho \sigma  }\partial _{\nu } A_{\lambda }\partial _{\rho } A_{\sigma }.\ee Since the translation symmetry in $x_4$ remains, $k_4$ is a good quantum number. Thus the above effective action is the sum of the Berry phase action of an assembly of 3D systems, each labeled
by a different $k_4$:
\be
&&\sum _{k_4} S_{3D}\left[k_4+\phi;A_{\mu }\right]\nn
&=&-i\frac{C_2\phi }{8 \pi ^2}L_4\int d^3x dt \epsilon ^{\nu \lambda \rho \sigma  }\partial _{\nu } A_{\lambda }\partial _{\rho } A_{\sigma }\label{sq}\ee
Subtract the $\phi=0$ version of \Eq{sq} from the above result, we obtain
\be&&\sum _{k_4} \left(S_{3D}\left[k_4+\phi;A_{\mu }\right]-S_{3D}\left[k_4;A_{\mu }\right]\right)\nn&=&-i\frac{C_2\phi }{8 \pi ^2}L_4\int d^3x dt \epsilon ^{\nu \lambda \rho \sigma  }\partial _{\nu } A_{\lambda }\partial _{\rho } A_{\sigma }.\label{diff}\ee

For non-compact gauge field, since the integrand of the right hand side is a pure derivative, it is only
nonzero in the presence of boundary. With spatial boundaries  \be &&-i\frac{C_2\phi }{8 \pi ^2}L_4\int d^3x dt \epsilon ^{\nu \lambda \rho \sigma  }\partial _{\nu } A_{\lambda }\partial _{\rho } A_{\sigma }\nn&&=-i\frac{C_2\phi }{8 \pi ^2}L_4\oint d^2x dt ~n_{\nu }[\epsilon ^{\nu \lambda \rho \sigma  } A_{\lambda }\partial _{\rho } A_{\sigma }]\label{scs}\ee
where $n_\nu$ is the $\nu$th component of the unit surface normal.  Note that the right hand
side of the above expression is the first Chern-Simons action defined on the surface suggesting surface  Hall effect as discussed in Ref. [\onlinecite{qi2008}]. Combining \Eq{diff} and \Eq{scs} we conclude
\be
\sum_{k_4}\Delta\s_{xy,{\rm ~surface}}(k_4)= {C_2\over 2\pi}{L_4\phi\over 2\pi}.\label{qqqq}\ee
Here $\Delta\s_{xy,{\rm ~surface}}(k_4)$ is the difference in the surface Hall conductance between the adiabatically evolved ($k_4\ra k_4+\phi$) system and the original ($k_4$) system. \Eq{qqqq} is in close analogy of \Eq{qqq} in earlier discussions.

Since  for two dimensional Dirac electrons a change  of mass from negative to positive induces  $\Delta \s_{xy}=1/2\pi$, the above result strongly suggests that as $k_4\ra k_4+\phi$, $C_2$ number of massless surface Dirac Hamiltonians are encountered. Let us label the $k_4$ values at which the surface Dirac Hamiltonian is massless by $k_4^*$s.  As these special $k_4^*$s are crossed under $k_4\ra k_4+\phi$, the mass of the surface Hamiltonians change sign. \Eq{qqqq} requires when we compare the masses of the adiabatic evolved surface Hamiltonians with the original ones we will find $C_2\times L_4\phi/2\pi$ mass reversals.
\begin{figure}[tbp]
\begin{center}
\includegraphics[scale=0.3]
{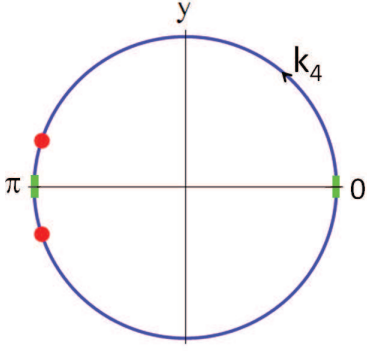}
\caption{(color on-line) The circle represents $k_4\in [0,2\pi)$
  and the red points mark $k_{4,1}^*$ and $k_{4,2}^*$. The green arcs in the vicinity of $k_4=0$ and $k_4=\pi$ mark the regime in $k_4$ where the surface bands merge into the bulk continuum. In these $k_4$ intervals the surface states on the top and bottom surfaces re-hybridize, and the surface Hall conductance vanishes.}
\label{point}
\end{center}
\end{figure}

\begin{figure}[tbp]
\begin{center}
\includegraphics[scale=0.3]
{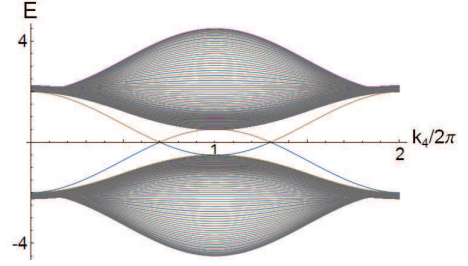}
\caption{(color on-line) The spectrum of \Eq{4dh} at surface momenta $k_1=k_2=0$ as a function of $k_4$. The two points at which the surface bandgap closes are $k_{4,1}^*$ and $k_{4,2}^*$.
.}
\label{k4}
\end{center}
\end{figure}

For the model discussed in \Eq{4dh} this is indeed true. Under the condition $x_3$ direction is open, while $x_{1,2}$ directions have periodic boundary conditions, the surface Hamiltonians read
\be
 &&H_{\rm surf}(\v k)\nn&&=\left\{\begin{array}{cc}
v~(-k_1\s_2+k_2\s_1)+ m(k_4)\s_3&{\rm top~surface}\\
v~(k_1\s_2+k_2\s_1)+ m(k_4)\s_3&{\rm bottom~surface}\end{array}\right.\nn&&
\label{srh}\ee
where $v$ is a velocity parameter, $k_{1,2}$ are the components of the surface wavevector. The mass term $m(k_4)$ vanishes when $k_4=k_4^*$. Surprisingly, despite the fact that the Hamiltonian in \Eq{4dh} is time-reversal invariant, the $k_4^*$s are not $0$ or $\pi$. Instead, we find a pair of $k_{4,1}^*$ and $k_{4,2}^*=2\pi-k_{4,1}^*$ where $m(k_4)$ vanishes. The value of these two $k_4^*$s value depend on the parameter $m$ in \Eq{dst}. For example for $m=-2.5$ we find $k_{4,1}^*=0.6992282\pi$
  and $k_{4,2}^*=2\pi-0.6992282\pi.$ In \Fig{k4} we show the $k_4$-dependence of the energy spectrum of $H$ in \Eq{4dh} under open boundary condition in $x_3$, and with $(k_1,k_2)$ fixed at $(0,0)$.  In addition to the gapless points $k_{4,1}^*$ and $k_{4,2}^*$ there is another important feature worthy of attention. In the vicinity of $k_4=0$ and $k_4=\pi$ (marked by green) the surface bands merge into the bulk continuum. As a result, the surface states on the top and bottom surfaces re-hybridize, resulting in the vanishing of the  surface Hall conductance. This is illustrated in \Fig{point} where the circle represents $k_4\in [0,2\pi)$
and the red points mark $k_{4,1}^*$ and $k_{4,2}^*$. The green arcs represent the intervals of $k_4$ in which the surface bands merge into the bulk continuum. It is important to note that there is no gap closing at either ends of these arcs.

\begin{figure}[tbp]
\begin{center}
\includegraphics[scale=0.3]{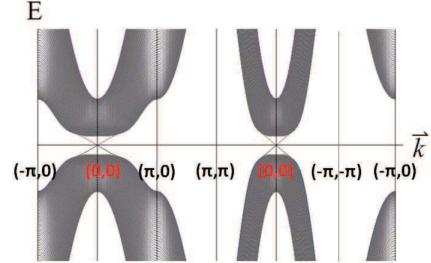}\vspace{-0.5in}
\caption{(color on-line) The surface band structure obtained by setting $k_4$ in \Eq{4dh} to one of the special values
(marked by the red dots) in \Fig{point}. }
\label{disp}
\end{center}
\end{figure}

In \Fig{disp}, we plot the energy spectrum associated with $k_{4,1}^*$ as a function of $k_{1,2}$. Note that despite the fact that for $k_4\ne 0$ or $\pi$ time-reversal is broken, the surface Dirac point is located at $\v k=(0,0)$. In addition, at $k_4=k_{4,1}^*$ and $k_{4,2}^*$ the low energy surface Hamiltonians are completely time-reversal invariant! The fact there exists two $k_{4}^*$s at which $k_4\ra k_4+\phi$ induces $m(k_4)$ to change from negative to positive, is consistent with the fact that $C_2=2$ in \Eq{qqqq}. Because of the $k_4^*$s are not time reversal invariant, there is no time reversal symmetry that protects the gapless Dirac surface states. However, finding symmetry protected gapless Dirac surface states is not the purpose of this paper.

\subsection{The WZW action in 1+1 D}
\label{wzw1}

The purpose of this section is to derive the NL$\s$ model describing the competing dimerization and magnetic order described at the beginning of section IV.

We start with the 4D topological insulator described by \Eq{4dh} and \Eq{dst} and insert magnetic fluxes $L_{1,2,3}\theta_{1,2,3}$ in the first to the third holes of the 4D torus. Moreover, we shall restrict $\theta_{1,2,3}$ to depend only on $x_4$ and $t$. By substituting $A_{1,2,3}\ra \theta_{1,2,3}(x_4,t)$ and fix the gauge so that $A_{0,4}\ra 0$, \Eq{cs2} becomes
\be
S_{CS_2}\ra -i{C_2\over 12\pi^2}L_1L_2L_3\int dt dx_4\e^{abc}\theta_a\p_0\theta_b\p_4\theta_c,\label{cs22}\ee
where $a,b,c=1,2,3$.
Since $k_{1,2,3}$ are good quantum numbers, \Eq{cs22} is the sum of the Berry phase action of
an assembly of one dimensional systems, each parameterized by $(k_1,k_2,k_3)$, i.e.,
\be &&\sum_{k_1,k_2,k_3}S_{1D}[k_{1,2,3}+\theta_{1,2,3}]\nn&&=-i{C_2\over 12\pi^2}L_1L_2L_3\int dt dx_4\e^{abc}\theta_a\p_0\theta_b\p_4\theta_c\nn&&\label{xs}\ee

Again, we introduce an auxiliary coordinate $u\in [0,1]$ so that $\theta_{1,2,3}(t,u)$ smoothly interpolate between
$\theta_{1,2,3}(t,u=0)=0$ and $\theta_{1,2,3}(t,u=1)=\theta_{1,2,3}(t)$, and consider the following action
\be -i{C_2\over 24\pi^2}\prod_{i=1}^3L_i\int dt dx_4 \int_0^1 du\e^{abc}\e^{\mu\nu\lambda}\p_\mu\Big(\theta_a\p_\nu\theta_b \p_\lambda\theta_c\Big)\nn
=-i{C_2\over 24\pi^2}\prod_{i=1}^3L_i\int dt dx_4 \int_0^1 du\e^{abc}\e^{\mu\nu\lambda} \p_\mu\theta_a\p_\nu\theta_b\p_\lambda\theta_c,\nn\label{wwt} \ee
where $\mu=0,1,2$ labels the $t, x_4$ and $u$ coordinates, respectively.  Because the integrand of \Eq{wwt} is a total derivative it is equal to a boundary integral. Since  boundaries only exist in the $u$ direction we can easily show that \Eq{wwt} reduces to \Eq{cs22}. As the result \Eq{xs} implies
\be  &&\int\frac{dk_1dk_2dk_3}{(2\pi)^3}S_{1D}[k_{1,2,3}+\theta_{1,2,3}] \nn
&=&-i{C_2\over 24\pi^2}\int dt dx_4 \int_0^1 du\e^{abc}\e^{\mu\nu\lambda}\p_\mu\theta_a\p_\nu\theta_b \p_\lambda\theta_c.\label{xxs}\ee

If we use $j$ to label the lattice sites in the $x_4$ direction, the 1D Hamiltonian parameterized by $k_{1,2,3}$
is given by
\be
&&H=-\sum_{j}\psi^\dagger_{j}\Big[(1-m_4^2)\Gamma_5 +\sum_{\alpha=1}^3 m_\alpha\Gamma_\alpha\Big]\psi_{j} \nn
&&-\sum_{j}\bigg[\psi^\dagger_{j+1}\Big[{1+m_4^2\over 2}\Gamma_5-i m_4\Gamma_4\Big]\psi_{j}+ h.c.\bigg],
\label{smh}\ee
where $m_{1,2,3,4}$ are given by
 \be
 &&m_1=\text{sin}(k_1+\theta_1), \nn&& m_2=\text{sin}(k_2+\theta_2),\nn
&&m_3=\text{sin}(k_3+\theta_3),\nn
&&m_4=\left(m+\text{cos}(k_1+\theta_1) +\text{cos}(k_2+\theta_2) +\text{cos}(k_3+\theta_3)\right).\nn
\ee In the following we shall set the parameter $m$ to $-2.5$.
 For $m_{1,2,3,4}=0$ the dispersion of the one dimensional Hamiltonian is shown in \Fig{ladder}(c) where both bands are doubly degenerate. The question is what kind of Berry phase action for $m_\alpha$'s  will be generated upon integrating out the fermions. (Due to the quadratic touching between the conduction and valence bands, the gradient expansion method discussed in Ref.[\onlinecite{abanov2000}] no longer works.)

It can be checked that  the following three-torus to $S^3$ mapping $(k_1,k_2,k_3)\ra n_\alpha(k_{1,2,3}+\theta_{1,2,3})\equiv m_\alpha (k_{1,2,3}+\theta_{1,2,3})/\sqrt{\sum_{\alpha=1}^4 m_\alpha^2(k_{1,2,3}+\theta_{1,2,3})}$ has Pontryagin index $P=1$ regardless of the $\theta_{1,2,3}$ values, where
 \be
P= {1\over 2\pi^2}\int_0^{2\pi}\prod_{\alpha=1}^3 dk_\alpha \e^{abcd} n_a\p_{k_1}n_b\p_{k_2}n_c\p_{k_3}n_d.\ee
Given this fact and \Eq{wwt}, we now show that the desired Berry phase action generated by integrating out the fermion is the WZW action
\be
 &&S_{WZW}[\hat{n}(x_4,t)]\nn
&=&-{i C_2\over 6\pi}\int_0^1 du \int dx_4dt \e^{\mu\nu\lambda}\e^{abcd} n_a\p_\mu n_b\p_\nu n_c\p_\lambda n_d.\nn&&\label{wt}\ee
The proof is similar to what we did earlier in section III, namely, just using chain rule of differentiation.
If integrating out fermion indeed generate \Eq{wt},  consistency with \Eq{xxs} would require
\be
&&\!\!\!\!\!\!\!{-i C_2\over 6\pi}\int \prod_{i=1}^3\frac{dk_i}{2\pi}\int_0^1 du \int dx_4dt \e^{\mu\nu\lambda}\e^{abcd} n_a\p_\mu n_b\p_\nu n_c\p_\lambda n_d\nn
&&=-i{C_2\over 24\pi^2}\int dt dx_4 \int_0^1 du\e^{abc}\e^{\mu\nu\lambda}\p_\mu\theta_a\p_\nu\theta_b \p_\lambda\theta_c.\label{pf}\ee
Since each of the $n_\alpha$ depends on $k_{1,2,3}+\theta_{1,2,3}(t,x_4,u)$ we apply chain rule to the left hand side of \Eq{pf}:
\begin{widetext}
\begin{eqnarray}
&&{-i C_2\over 6\pi}\int {dk_1dk_2dk_3\over (2\pi)^3}\int_0^1 du \int dx_4dt \e^{\mu\nu\lambda}\e^{abcd} n_a\p_\mu n_b\p_\nu n_c\p_\lambda n_d\nn
&&={-i C_2\over 6\pi}\int {dk_1dk_2dk_3\over (2\pi)^3}\int_0^1 du \int dx_4dt \e^{\mu\nu\lambda}\e^{abcd} n_a\p_{\theta_\alpha} n_b\p_{\theta_\beta} n_c\p_{\theta_\gamma} n_d \left(\p_\nu\theta_\alpha\p_\nu\theta_\beta\p_\lambda\theta_\gamma\right)\nn
&&={-i C_2\over 6\pi}\int {dk_1dk_2dk_3\over (2\pi)^3}\int_0^1 du \int dx_4dt \e^{\mu\nu\lambda}\e^{abcd} n_a\p_{\theta_1} n_b\p_{\theta_2} n_c\p_{\theta_3} n_d \e^{\alpha\beta\gamma}\left(\p_\nu\theta_\alpha\p_\nu\theta_\beta\p_\lambda\theta_\gamma\right)\nn
&&={-i C_2\over 48\pi^4}\int_0^1 du \int dx_4dt \e^{\mu\nu\lambda}\e^{\alpha\beta\gamma}\Big(\int dk_1dk_2dk_3\e^{abcd}n_a\p_{k_1} n_b\p_{k_2} n_c\p_{k_3} n_d \Big) \left(\p_\nu\theta_\alpha\p_\nu\theta_\beta\p_\lambda\theta_\gamma\right)\nn&&=
{-i C_2\over 24\pi^2}\int_0^1 du \int dx_4dt \e^{\mu\nu\lambda}\e^{\alpha\beta\gamma} \left(\p_\nu\theta_\alpha\p_\nu\theta_\beta\p_\lambda\theta_\gamma\right).
\end{eqnarray}
\end{widetext}
Indeed, it agrees with the right hand side of \Eq{pf}!

The non-linear $\sigma$ model with the $C_2=2$ WZW term
\be
S=\int dx_4 dt \p_\mu n_\alpha  \p_\mu n_\alpha +S_{WZW}[n_\alpha(x_4,t)]\label{wzww}
\ee
describes the critical point between the dimerised phase ($\langle m_0\rangle \ne 0$) and the Haldane phase. This critical theory is realized in the spin-1 chain with quadratic and bi-quadratic exchanges:
\be
H=J \sum_i\left[\v S_i\cdot\v S_{i+1}-(\v S_i\cdot\v S_{i+1})^2\right].
\ee
It has been exactly solved by Takhtajan\cite{takhtajan1982} and Babujan\cite{babujan1983}, and the critical theory is that of level-2 $SU(2)$ WZW model whose NL$\s$ model form is given by \Eq{wzww}.
\\

\subsection{The $S=1$ spin chain, edge states, and the $\theta$ term in the NL$\s$ model}
\begin{figure}[tbp]
\begin{center}
\includegraphics[scale=0.3]
{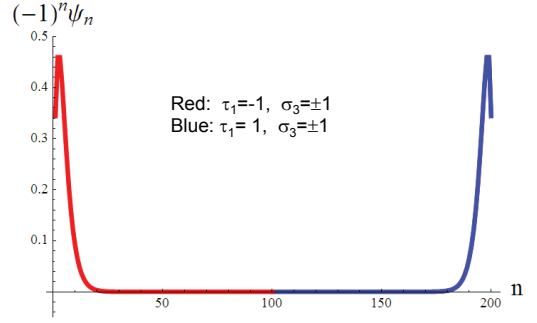}
\caption{(color on-line) The envelope of the edge wavefunctions.}
\label{edge}
\end{center}
\end{figure}

In this section we set $ m_{1,2,3}\ra\cos\theta_0\hat{m}_{1,2,3}$ and $m_4\ra\sin\theta_0$. We
consider the following fermion $\s$ model
\be
H&=&-\sum_{j}\psi^\dagger_{j} \Big[(1-\sin^2\theta_0)\Gamma_5+\cos\theta_0\sum_{k=1}^3 \hat{m}_k \Gamma_k\Big]\psi_{j}\nn &-&\sum_{j}\bigg[\psi^\dagger_{j+1}\Big[{1+\sin^2\theta_0\over 2}\Gamma_5-i\sin\theta_0\Gamma_4\Big]\psi_{j}+ h.c.\bigg],\nn
\label{chain}\ee
where the dimerization order parameter has been set to $\sin\theta_0$ and the magnetic order parameters to $\cos\theta_0 \hat{m}$. In the absence of the $m_k$'s \Eq{chain} describes a one-dimensional TBI whose bandstructure is shown in Fig.2(a) (each band is now two-fold degenerate). Under the open boundary condition there is a doubly degenerate zero-energy edge states at each end of the chain.
Using the representation
\be
\Gamma_{1,2,3}=\tau_1\otimes\s_{1,2,3},~\Gamma_4=\tau_2\otimes I,~\Gamma_5=\tau_3\otimes I,\ee
the edge states can be made eigenstates of $\tau_1$ and $\s_3$. The wavefunctions of these edge states are shown in
\Fig{edge}. Notice that we have plotted $(-1)^n\psi_n$ ($n$ is the site index) rather than simply $\psi_n$. This is because the minimum gap occurs at $k=\pi$ hence the these wavefunctions have a fast spatial modulation.
In the presence of non-zero $\hat{m}=(\hat m_1,\hat m_2,\hat m_3)$ the degenerate edge states are split. The occupied state has magnetization along the local $\hat{m}$ direction. In the presence of space-time dependent $\hat{m}$, integrating out the fermions produces space-time dependent edge magnetic moments each contributing a Berry phase of the form given by \Eq{wzs}. This is precisely the well-known spin-1/2 at the ends of the spin-1 chain.

By letting $\hat{n}$ in \Eq{wt} smoothly interpolate between $\hat{n}(t,x_4,u=1)=(0,0,0,1)$ and $\hat{n}(t,x_4,u=0)=\left(\sin\theta_0, \cos\theta_0\hat{m}(x_4,t)\right)$ we obtain the $\theta$-action for gapped spin chain
\begin{eqnarray}
S_\theta=-i \theta\left[\frac{1}{4\pi}\int dx_4 dt \hat{m}\cdot\p_t\hat{m}\times\p_{x_4}\hat{m}\right],
\end{eqnarray}
where $\theta = C_2[\pi -\sin(2 \theta _0)-2 \theta _0]$. When the system has some sort of discrete symmetry, such as inversion, which requires $m_4=0$ or $\theta_0=0$, the coefficient $\theta=C_2\pi$. If $C_2$ is an odd integer, the corresponding NL-$\s$ model is gapless; if $C_2$ is an even integer, the NL$\s$ model is gapped.
When there is no such discrete symmetry, the bare coefficient of the $\theta$ term, in general, deviates from odd integral multiple of $\pi$, at low energy it renormalizes to the closest fixed point characterized by the $\theta$ term coefficient $2 n\pi$, which corresponds to a gapped system. In our particular case the relevant value is $2\pi$. This is in analogy with the $Z_2$ classification of 3D TBIs and the discrete symmetry, such as inversion, which protect the gapless modes for $\theta=(2n+1)\pi$, plays the role of time reversal symmetry in 3D TBIs.
\\
\section{Summary}

In this paper we have tackled the issue of what is the relation between the following two areas of topological condensed matter physics: topological band insulators and topological non-linear $\s$ (NL$\s$) models. We have established a correspondence between a $2n$-dimensional Chern-Simons band insulators and the $(n-1)$-dimensional NL$\s$ models with the Wess-Zumino-Witten (WZW) term. The correspondence does not depend on whether in the absence of the order parameter (the $\s$ fields) the original gapless $n-1$ dimensional fermion Hamiltonian is Dirac-like or not. We have applied this correspondence to a simple 1D model where the gapless Hamiltonian exhibits a quadratic band touching. In the presence of the dimerization and antiferromagnetic order parameters, the fermion integration yields an $O(4)$ non-linear $\s$ model whose WZW term has a doubled coefficient. This action describes the  critical point between the dimerized phase and the Haldane phase. We demonstrate that the the edge states of the $S=1$ spin chain are nicely captured if one starts with the edge state of the dimerized 1D topological band insulator.

\acknowledgments We thank Ying Ran, Ashvin Vishwanath, Tao Xiang, and Guang-Ming Zhang for useful discussions. This work is supported by DOE grant number DE-AC02-05CH11231.

\appendix
\begin{widetext}
\section{The proof of the correspondence between a $2n$-dimensional Chern-Simons insulator and a $(n-1)$-dimensional
topological NL$\s$ model}
\label{proof}
In this appendix we provide the  proof of the Chern-Simons insulators $\leftrightarrow$ topological NL$\s$ model correspondence
discussed in the introduction.

We consider a $(n-1)$-dimensional fermion-$\s$ model described by the time-dependent Hamiltonian
$H(\vec k;\vec m(t,\vec x))$, where $\vec x=(x_1,\cdots,x_{n-1})$ and $\vec k=(k_1,\cdots,k_{n-1})$ are the position and momentum in $(n-1)$-dimensions, and $\vec m(t,\vec x)$ is the space-time dependent $(n+2)$-component order-parameter fields with a fixed $|\vec m(t,\vec x)|$. We assume that the space and time derivatives of $\vec m(t,\vec x)$ do not enter the Hamiltonian.
When the order-parameter fields $\vec m(t,\vec x)$ are set to zero, the corresponding Hamiltonian $H_0(\vec k)=H(\vec k;\vec m(t,\vec x)\ra 0)$ is gapless. One familiar case is when $H_0(\vec k)$ has linear dispersion at low energy. However, our following results apply for more general situations.

The condition we need to impose to $H_0(\vec k)$ is that for constant $m_a$ the Hamiltonian $H(\vec k;\vec m)$
has a completely gapped spectrum, and  for $0<|\vec m|<m^\ast$ the energy gap does not close. The conditions we impose above apply for nearly all physically interesting situations. (Note that, in a continuum model where the linear Dirac dispersion extends to $E=\pm\infty$, $m^\ast$ can be taken to $\infty$ since the energy gap never closes for any non-zero $|\vec{m}|$. This may not be true for lattice fermion models where the bandwidth of $H_0(\vec k)$ is finite.) Physically $m^\ast$ should be of the order of the band width of $H_0(\vec k)$.
For such $H(\vec k;\vec m(t,\vec x))$, we take the following steps to determine whether a WZW term will be generated by fermion integration.

Step 1, we create a $2n$-dimensional Bloch Hamiltonian $ H(\vec k;\lambda\vec m(\vec p))$, where $\vec p=(p_1,\cdots,p_{n+1})$ and $(\vec k,\vec p)$ is the $2n$-component momentum, by simply replacing constant $m_a$ in $H(\vec k;\vec m)$ with $\lambda m_a(\vec p)$. Most importantly we choose $\vec m(\vec p)$ such that it has a nonzero Pontryagin index $P$ (\Eq{pont}). For instance, the following map
\be
&&m_i(\vec p)=\sin p_i, {\rm ~for~}i=1,\cdots, n+1,\nn
&&m_{n+2}(\vec p)=(m-\sum_{i=1}^{n+1} \cos p_i)
\ee
has the Pontryagin index (\Eq{pont}) equal to one when the parameter $m$ satisfies $n-1<m<n+1$. Note that the parameter $\lambda$ is introduced so that  $|\lambda\vec m(\vec p)|<m^\ast$.
For such $|\lambda\vec m(\vec p)|$, the Hamiltonian $H(\vec k;\lambda\vec m(\vec p))$ describes an insulator so that we can go step 2 below.

Step 2, we check whether the $2n$-dimensional Bloch Hamiltonian $ H(\vec k;\lambda\vec m(\vec p))$ possesses a non-zero $n$-th Chern number $C_n$ (\Eq{cn}). If $C_n\neq 0$, a level-$k=C_n/P$  WZW term (\Eq{hut}) will be generated upon integrating out fermions in the $(n-1)$-dimensional Hamiltonian $H(\vec k; \vec m(t,\vec x))$, so long as $|\vec{m}(t,\vec{x})|<m^\ast$.
The following is the proof.

The electromagnetic gauge action of a $2n$-dimensional Chern-Simons insulator contains the following topological term\cite{golterman1993,qi2008}
\be
S_{CS_n}
=-i\frac{C_n}{(n+1)!(2\pi)^n}\int dt d^{2n}x \Big[\epsilon^{a_1\cdots a_{2n+1}} A_{a_1}\p_{a_2} A_{a_3}\cdots \p_{a_{2n}} A_{a_{2n+1}}\Big],\nn
\label{csn}
\ee
where $a_i=0,1,\cdots,2n$ labels $t$, $x_1,\cdots,x_{2n}$ respectively, $\epsilon$ is the $(2n+1)$d antisymmetric tensor. Now, we choose
\be
&&A_\mu(t,x_1,\cdots,x_{2n})=0,~\textrm{for}~\mu=0,\cdots,n-1;\nn
&&A_{i+n-1}(t,x_1,\cdots,x_{2n})=\theta_i (t,x_1,\cdots,x_{n-1}),~\textrm{for}~i=1,\cdots,n+1.
\ee
For this choice of gauge potential, the Chern-Simons action becomes
\be
S_{CS_n}\to
-i\frac{C_n}{(n+1)!(2\pi)^n}\left[\prod_{i=n}^{2n}L_i\right]\int dtd^{n-1}x~ \epsilon^{a_1\cdots a_{n}}\epsilon^{b_1\cdots b_{n+1}} \theta_{b_1} \p_{a_1} \theta_{b_2}\cdots \p_{a_{n}} \theta_{b_{n+1}}
\label{snp}\ee
where $a_i=0,1,\cdots,n-1$ represents $t,x_1,\cdots,x_{n-1}$ respectively, $b_i=1,\cdots,n+1$, and
$d^{n-1}x\equiv dx_1\cdots dx_{n-1}$. (Note that a overall $n$-dependent sign is ignored above and hereafter since it  can be absorbed by relabeling the components of $\hat{n}$.) We introduce an auxiliary coordinate $u\in [0,1]$ so that $\theta_i(t,\vec x,u)$ ($i=1,\cdots,n+1)$ smoothly interpolate between $\theta_i(t,\vec x,u=0)=0$ and $\theta_i(t,\vec x,u=1)=\theta_i(t,\vec x)$. [$\vec{x}\equiv (x_1,\cdots,x_{n-1})]$. In the following $a_i=0,1,\cdots,n$ will correspond to $t,x_1,\cdots,x_{n-1},u$ respectively, and $\epsilon^{01\cdots n}=1$. Under this condition \Eq{snp}  can be rewritten as
\be
S_{CS_n}
&=&-i\frac{C_n}{(n+1)!(2\pi)^n}\left[\prod_{i=n}^{2n}L_i\right]\int dtd^{n-1}x\int_0^1 du~ \epsilon^{a_1\cdots a_{n}a_{n+1}}\epsilon^{b_1\cdots b_{n+1}}\p_{a_{n+1}}\Big[ \theta_{b_1} \p_{a_1} \theta_{b_2}\cdots \p_{a_{n}} \theta_{b_{n+1}}\Big],\label{A5}\\
&=&-i\frac{C_n}{(n+1)!(2\pi)^n}\left[\prod_{i=n}^{2n}L_i\right]\int dtd^{n-1}x\int_0^1 du~ \epsilon^{a_1\cdots a_{n}a_{n+1}}\epsilon^{b_1\cdots b_{n+1}}\Big[\p_{a_{n+1}} \theta_{b_1} \p_{a_1} \theta_{b_2}\cdots \p_{a_{n}} \theta_{b_{n+1}}\Big],\label{A6}\\
&=&-i\frac{C_n}{(n+1)!(2\pi)^n}\left[\prod_{i=n}^{2n}L_i\right]\int dtd^{n-1}x\int_0^1 du~ \epsilon^{a_1\cdots a_{n}a_{n+1}}\epsilon^{b_1\cdots b_{n+1}} \Big[\p_{a_{1}}\theta_{b_1} \p_{a_2} \theta_{b_2}\cdots \p_{a_{n+1}} \theta_{b_{n+1}}\Big].\label{A7}
\ee
In passing from \Eq{snp} to \Eq{A5}, we have used the fact that the system is periodic in $t, x_1,..,x_{n-1}$ directions; the integral in \Eq{A5} is equal to \Eq{snp} because only the $u$-direction has boundaries.

Since the $\theta$'s do not depend on the coordinates $(x_n,\cdots,x_{2n})$, translation symmetry in these $(n+1)$ directions remain intact, and the associated momenta $\vec{p}$ are good quantum numbers. As a result the $2n$-dimensional system decouples into $\left[\prod_{i=n}^{2n}L_i \right]$ number of  $(n-1)$-dimensional subsystems each labeled by a different $\vec p$, and is described by the Hamiltonian $\tilde H(\vec k;\vec m(\vec p+\vec \theta(t,\vec x)))$. Let $S_{(n-1)D}(\vec p)$ be the topological action of each of these $n-1$ dimensional systems (upon integrating out the fermions), they satisfy
\be
\sum_{\vec p} S_{(n-1)D}(\vec p)=S_{CS_{n}}.
\label{relation}
\ee
For a fixed $\vec{\theta}(t,\vec x)$ profile, the Pontryagin index of the map $\vec p\to\hat n(\vec p+\vec \theta)\equiv  \vec m(\vec p+\vec \theta)/|\vec m(\vec p+\vec \theta)|$ is given by
\be
P&=&\frac{1}{\textrm{Area}(S^{n+1})}\int d^{n+1}\vec p~ \epsilon^{c_1\cdots c_{n+2}}n_{c_1}(\vec p+\vec\theta)\p_{p_1} n_{c_2}(\vec p+\vec\theta)\cdots \p_{p_{n+1}} n_{c_{n+2}}(\vec p+\vec \theta),\\
&=&\frac{1}{\textrm{Area}(S^{n+1})}\int d^{n+1}\vec p~ \epsilon^{c_1\cdots c_{n+2}} \big[n_{c_1}(\vec p+\vec \theta)\p_{\theta_1} n_{c_2}(\vec p+\vec \theta)\cdots \p_{\theta_{n+1}} n_{c_{n+2}}(\vec p+\vec \theta)\big]\Big|_{\vec\theta\to\vec{\theta}(t,\vec x)}.
\ee
It is important to note that the Pontryagin index of $\vec m(\vec p+\vec \theta)$ is independent on $\vec\theta$. Now, we can rewrite $S_{CS_n}$ as follows:
\be
S_{CS_n}&=&S_{CS_n}\cdot\left[\frac{1}{P} \frac{1}{\textrm{Area}(S^{n+1})}\int d^{n+1}\vec p~ \epsilon^{c_1\cdots c_{n+2}}n_{c_1}\p_{\theta_1} n_{c_2}\cdots \p_{\theta_{n+1}}n_{c_{n+2}} \right]
\\
&=&-i\frac{1}{(n+1)!(2\pi)^n} \frac{C_n/P}{\textrm{Area}(S^{n+1})}\left[\prod_{i=n}^{2n} L_i\right]\int d^{n+1}\vec p \int dt d^{n-1} \vec x\int_0^1 du \Big[ \epsilon^{c_1\cdots c_{n+2}}n_{c_1}\p_{\theta_1} n_{c_2}\cdots \p_{\theta_{n+1}}n_{c_{n+2}} \Big] \nn
&&\qquad\qquad\qquad\qquad\qquad\qquad\qquad\qquad\qquad  \qquad\times\Big[\epsilon^{a_1\cdots a_{n+1}} \epsilon^{b_1\cdots b_{n+1}}\p_{a_{1}}\theta_{b_1} \p_{a_2} \theta_{b_2}\cdots \p_{a_{n+1}} \theta_{b_{n+1}}\Big],\label{A14}\\
&=&\sum_{\vec p} \left[-i{C_n\over P}\frac{2\pi}{\textrm{Area}(S^{n+1})(n+1)!}  \int dt d^{n-1} \vec x\int_0^1 du~  \epsilon^{c_1\cdots c_{n+2}}\epsilon^{a_1\cdots a_{n+1}} n_{c_1}\p_{a_1} n_{c_2}\cdots \p_{a_{n+1}}n_{c_{n+2}} \right],\label{A15}
\ee
In going from \Eq{A14} to \Eq{A15} we have applied the chain rule of differentiation. Note that because, for different $\vec p$, the Hamiltonians $H(\vec k;\lambda\vec m(\vec p+\vec \theta(t,\vec x)))$ can be adiabatically deformed into one another without closing the energy gap, the subsystems correspond to different $\vec{p}$ should share the same form of WZW term with the same quantized coefficient.
Since $S_{CS_n}= \sum_{\vec{p}} S_{(n-1)D}(\vec p)$, we conclude that
\be
S_{(n-1)D}(\vec p)&&=-i\frac{C_n}{P} \frac{2\pi}{\textrm{Area}(S^{n+1})(n+1)!}  \int dt d^{n-1}  x\int_0^1 du~  \epsilon^{c_1\cdots c_{n+2}}\epsilon^{a_1\cdots a_{n+1}} n_{c_1}\p_{a_1} n_{c_2}\cdots \p_{a_{n+1}}n_{c_{n+2}}\nn&&=-i\frac{C_n}{P} \frac{2\pi}{\textrm{Area}(S^{n+1})}  \int dt d^{n-1}  x\int_0^1 du~  \epsilon^{c_1\cdots c_{n+2}}n_{c_1}\p_{t} n_{c_2}\cdots \p_{u}n_{c_{n+2}}
\label{des}
\ee
We note that in \Eq{des} the dependence on $\vec{p}$ is implicitly contained
in  $\hat n(\vec p+\vec \theta(t,\vec{x}))$. By replacing $\vec m(\vec p+\vec \theta(t,x,u))$ with $\vec m(t,\vec x,u)$, which smoothly interpolates between $\vec m(t,\vec x,u=0)=\vec m(\vec p)$ and $\vec m(t,\vec x,u=1)=\vec m(\vec p+\vec\theta(t,x))$, we obtain the following topological action upon integrating out the fermions in \Eq{ag} with $\hat n(t,\vec x,u)=\vec m(t,\vec x,u)/|\vec m(t,\vec x,u)|$
\be
S_{(n-1)D}=-i\frac{C_n}{P} \frac{2\pi}{\textrm{Area}(S^{n+1})}  \int dt d^{n-1}  x\int_0^1 du~  \epsilon^{c_1\cdots c_{n+2}} n_{c_1}\p_{t} n_{c_2}\cdots \p_{u}n_{c_{n+2}},
\label{des1}
\ee
which is exactly the WZW term we desire!
\end{widetext}


\end{document}